# Spectra of the $D_2O$ dimer in the O-D fundamental stretch region: vibrational dependence of tunneling splittings and lifetimes

A.J. Barclay,[1] A.R.W. McKellar,[2] and N. Moazzen-Ahmadi[1]

[1] *Department of Physics and Astronomy, University of Calgary, 2500 University Drive North West, Calgary, Alberta T2N 1N4, Canada*

[2]*National Research Council of Canada, Ottawa, Ontario K1A 0R6, Canada*

**Abstract**

The fundamental O-D stretch region (2600 – 2800 $cm^{-1}$) of the fully deuterated water dimer, $(D_2O)_2$, is studied using a pulsed supersonic slit jet source and a tunable optical parametric oscillator source. Relatively high spectral resolution (0.002 $cm^{-1}$) enables all six dimer tunneling components to be observed, in most cases, for the acceptor asymmetric O-D stretch, the donor free O-D stretch, and the donor bound O-D stretch vibrations. The dominant acceptor switching tunneling splittings are observed to decrease moderately in the excited O-D stretch states, to roughly 75% of their ground state values, whereas the smaller donor-acceptor interchange splittings show more dramatic and irregular decreases. Excited state predissociation lifetimes, as determined from observed line broadening, show large variations ($0.2 \le \tau \le 5$ nanoseconds) depending on vibrational state, $K$-value, and tunneling symmetry. Another very weak band is tentatively assigned to a combination mode involving an intramolecular O-D stretch plus an intermolecular twist overtone. Asymmetric O-D stretch bands of the mixed isotopologue dimers $D_2O$-DOH and $D_2O$-HOD are also observed and analyzed.

## I. Introduction

The water dimer has been the subject of numerous studies using high-resolution spectroscopy and other techniques. Analyzing dimer spectra gives direct and precise information on the intermolecular forces between two water molecules which in turn provides a basis for better understanding of the behavior of liquid and solid water on a microscopic scale. Recent review articles give a detailed listing of many previous spectroscopic (and other) studies of water dimers.[1,2]

The present paper concerns infrared spectra of $(D_2O)_2$ in the region of the O-D stretch fundamentals (2600 – 2800 $cm^{-1}$). This region was previously studied by Paul et al.,[3,4] and our results constitute an extension of their work to significantly higher spectral resolution. High resolution enables dimer tunneling splittings to be better resolved, giving information on their vibrational dependence, and also enables dimer line widths to be measured, giving information on upper state predissociation lifetimes. In addition to these previously observed and relatively strong $(D_2O)_2$ fundamental bands, we also assign a weak band to a combination vibration involving the sum of an intramolecular O-D stretch mode plus a (low frequency) intermolecular mode. This is probably the first such mid-infrared observation of an intermolecular mode. Finally, we also analyze some bands of $D_2O$-DOH and $D_2O$-HOD, which again may be the first such mid-infrared observation of mixed water dimer isotopologues.

As well as the current and previous[3,4] studies of the O-D stretch fundamental, other pertinent high resolution mid-infrared water dimer spectroscopy has covered the monomer bending fundamental for both $(H_2O)_2$ and $(D_2O)_2$,[5,6] and the monomer O-H stretch fundamentals for $(H_2O)_2$.[7-9] Recently, there has also been activity concerning O-H stretching overtones in the near infrared region, revealing rich spectra that are not yet fully understood.[10-12]

## II. Background

The fascinating structure, symmetry effects, and tunneling dynamics of the water dimer have been described many times, but it is useful to summarize some of this material here with specific reference to $(D_2O)_2$, in order to explain the current results. Figure 1 illustrates the equilibrium dimer structure. One water monomer acts as a proton (or deuteron) donor and the other one as an acceptor. On the acceptor, the two D atoms are equivalent, while on the donor there are inequivalent “bound” and “free” D atoms (3 and 4, respectively, in Fig. 1). A plane of symmetry contains the two O atoms and the donor D atoms. The *a* inertial axis coincides approximately with the line connecting the O atoms, the *b*-axis is perpendicular to the symmetry plane, and the *c*-axis lies in the plane almost perpendicular to the O – O axis.

But the water dimer is by no means rigidly locked into this equilibrium structure -- in fact, it is highly nonrigid. The most facile nonrigid motion, giving rise to the largest tunneling splitting, involves interchange of the acceptor D atoms (1 and 2 in Fig. 1).The actual minimum energy interchange tunneling path is not a simple rotation of the acceptor, but rather a concerted motion of both monomers which has been likened[13] to the internal rotation of the methyl amine molecule. This acceptor switching motion splits each rotational level into two sublevels, usually labeled “1s” (ones) and “2s” (twos).The resulting energy difference is not yet directly known from experiment because only sums or differences of the splitting in different $K_a$ states have been measured.[14-16] However, Paul et al.3 have argued persuasively that the splitting value for the ground ($J = K_a = 0$) state of $(D_2O)_2$ is about 53 GHz (1.76789 cm$^{-1}$), and this value is approximately supported by theory.[17] For convenience in the current paper we will assume the value of exactly 53.0 GHz. Tunneling splittings are naturally much larger for $(H_2O)_2$ than for $(D_2O)_2$.

The next most significant water dimer tunneling effect involves interchange of the roles of the donor and acceptor monomers. Each sublevel is thus further split into three, and the resulting splitting, which can be directly measured, turns out to have a value of 0.0391 $cm^{-1}$ in the ground state of $(D_2O)_2$ as shown in Fig. 2a. The resulting symmetries are $A_1$, E, and $B_1$ for the "1s" levels, and $A_2$, E, and $B_2$ for the "2s". For convenience, the E symmetry level associated with $A_1$ and $B_1$ is usually labeled $E_1$, and that associated with $A_2$ and $B_2$ is labeled $E_2$. The nuclear spin weights for each symmetry in $(D_2O)_2$ are shown in parentheses in Fig. 2a.

The third and least facile tunneling effect involves interchange of the donor D atoms, and is known as bifurcation tunneling. This causes no further level splitting, but only level shifts which are quite small (≈0.0005 $cm^{-1}$) for the ground state of $(D_2O)_2$. As a result of bifurcation tunneling the E levels are no longer exactly midway between their A and B partners. Water dimer vibrational modes can be either in-plane ($A'$) or out-of-plane ($A''$), and in the present work we study the $A'$ donor free and bound O-D stretch fundamentals, and the $A''$ acceptor asymmetric O-D stretch fundamental.

Labeling and keeping track of the $(D_2O)_2$ energy levels is somewhat complicated by the fact that acceptor switching ("1s" or "2s") symmetry flips depending on even or odd $K_a$-value, and depending on $A'$ or $A''$ vibrational state symmetry. As well, interchange (A or B) symmetry flips depending on even or odd $J$-value. In this paper, we use rotationless A/B labels, which have to be multiplied by the appropriate rovibrational symmetries to give the full symmetry label. The labels in Fig. 2a are appropriate for rotational levels with ($K_a$, $K_c$) = (e, e) and (o, o) (which have + parity for the "1s" levels, and – parity for the "2s"). Table I shows the connection between our labels and those used in Ref. 16 and elsewhere; for example, the $A_1$ label is more fully expressed as $A_1^+/B_1^-$. Selection rules for rotational and rotation-vibration transitions are $A_1^+ \leftrightarrow A_1^-$, $B_1^+ \leftrightarrow B_1^-$, $A_2^+ \leftrightarrow A_2^-$, $B_2^+ \leftrightarrow B_2^-$. All transitions with $E^+ \leftrightarrow E^-$ are allowed in principle, but so far no transitions connecting the $E_1$ and $E_2$ manifolds have been detected.

The most complete analysis of the ground vibrational state of $(D_2O)_2$was made by Harker et al.,[16] who compiled all previous spectroscopic data[13-15,18-20] and added some new measurements. At present, the ground state levels with $K_a$ = 0, 1, and 2 are known for *J*-values up to about 10 or 11 for all six symmetry states. At the temperature of our supersonic jet (≈2 K), only ground state levels with $K_a$ = 0 and 1 and *J* less than about 7 are sufficiently populated to observe as initial states. Figure 2b shows the *origins* of these $K_a$ = 0 and 1 initial state levels, that is, the energies of the *J* = 0 levels (which of course are hypothetical for $K_a$ = 1). As mentioned, the ordering of "1s" and "2s" states for $K_a$ = 1 is reversed relative to $K_a$ = 0. On the scale of Fig. 2b, the interchange splittings are only barely visible. Considering $(D_2O)_2$ as an asymmetric rotor, the *A* rotational parameter is approximately equal to 4 cm$^{-1}$, but this is not particularly well defined because of the large and highly $K_a$-dependent acceptor tunneling splittings shown in Fig. 2b. The *B* and *C* parameters are each equal to about 0.18 cm$^{-1}$. The value of $(B - C)$ is only about 0.001 to 0.002 cm$^{-1}$, meaning that asymmetry doubling is quite small, even for $K_a$ = 1.

In the current paper, we use the parameters of Ref. 16 to calculate and fix the ground vibrational state energies of $(D_2O)_2$, and then fit excited state levels using the same empirical rotational energy expression,

$$E = \sigma + B_{av}[J(J+1) - K^2] - D[J(J+1) - K^2]^2 \pm [(B - C)/4][J(J+1)].$$

In this expression, $K = K_a$, and $B_{av} = (B + C)/2$. Each interchange tunneling level (A, B, E) of each acceptor switch level ("1s", "2s") of each $K_a$-value has its own origin, σ, and rotational parameters, $B_{av}$, $D$, $(B - C)$. The $(B - C)$ term was only used for $K = 1$ levels because we were not able to measure the small asymmetry doublings in excited $K = 2$ states, though they are known for the ground state.[15,16] As noted, each tunneling state and $K$-value is allowed to have its own rotational parameters. In contrast to this $K$ dependence, the tunneling splittings are assumed to be independent of $J$-value (a possible alternative might be to use common rotational parameters while allowing splitting to be $J$-dependent). The σ-values determined using this expression obviously depend on the presence or absence of the $- K^2$ terms: to avoid any ambiguity, our assumed ground state origins (derived from Ref. 16) are included in Table I.

**Results**

Spectra were recorded at the University of Calgary as described previously,[21-23] using a pulsed supersonic slit jet expansion probed by a rapid-scan optical parametric oscillator source. The usual expansion mixture contained about 0.007% $D_2O$ in helium carrier gas with a backing pressure of about 12 atmospheres. Limited spectra were also recorded with about 0.025% $D_2O$ in argon with a backing pressure of 2 atmospheres. Wavenumber calibration was carried out by simultaneously recording signals from a fixed etalon and a reference gas cell containing $N_2O$ or $C_2H_2$. Spectral assignment and simulation were made using the PGOPHER software.[24]

**A. The acceptor asymmetric O-D stretch band**

This band has been observed previously by Paul et al.,[3] who used a Raman shifted pulsed dye laser source, a pulsed supersonic slit jet, and the cavity ring down absorption technique, obtaining an effective instrumental resolution of about 0.03 cm$^{-1}$. Our instrumental width is at least ten times smaller, allowing most individual interchange splittings to be at least partially resolved for the first time. The asymmetric stretch vibration of the acceptor $D_2O$ monomer gives rise to perpendicular ($b$-type; $\Delta K_a = \pm1$, $\Delta K_c = \pm1$) rotational selection rules for the dimer, and we observed the resulting $K_a = 1 \leftarrow 0$, $0 \leftarrow 1$, and $2 \leftarrow 1$ subbands. The $A''$ symmetry of the excited vibration means that the ordering of the acceptor switching levels ("1s" and "2s") is flipped relative to the ground state.

The strongest subband, with $K_a = 1 \leftarrow 0$, is shown in Fig. 3. We assign the prominent *Q*-branch features at 2786.43 and 2787.73 $cm^{-1}$ to the "2s" and "1s" acceptor tunneling states, respectively, in agreement with Ref. 3. Each line of the "1s" subband consists of three components, and these are easily assigned to $B_1$ (lower), $E_1$ (middle), and $A_1$ (higher wavenumber) based on the intensity alternation caused by the spin weights shown in Fig. 2 (recall that our $A_1$ stands for $A_1^+/B_1^-$, etc.). The lines of the "2s" subband consist of a prominent central component, assigned to $E_2$ and a weaker component about 0.02 $cm^{-1}$ higher, assigned to $A_2$. The remaining $B_2$ component is not so obvious at first sight, but careful examination reveals that there is a broad shoulder on each "2s" line located about 0.01 $cm^{-1}$ below the central $E_2$ component. It was thus evident that the $B_2$ lines must be broadened due to predissociation effects in the upper state, and indeed most other lines in this subband also showed measurable but smaller predissociation broadening. In order to estimate the broadening we used profile fits, as described in detail below. Using these profile fits plus line position fits, we obtained the parameters listed in Table II for the $K = 1$ state of the acceptor asymmetric stretch fundamental. Note that these excited state origins (and others in this paper) depend on the assumed lower state origins as given in Table I.

The weaker $K_a = 0 \leftarrow 1$ and $2 \leftarrow 1$ subbands are shown in Fig. 4. Our assignments of the "1s" and "2s" *Q*-branches again agree with those of Paul et al.[3] These subbands are of course considerably weaker than $K_a = 1 \leftarrow 0$ because of smaller thermal populations in the $K_a = 1$ lower states, so the signal-to-noise ratio is reduced and there are more gaps in the observed spectra due to interference from $D_2O$ monomer lines. The “2s” transitions tend to be relatively stronger here than in the $K_a = 1 \leftarrow 0$ subband since the energy difference between the $K_a$ = 0 and 1 levels is smaller for “2s” than for “1s” (see Fig. 2b). In the $K_a = 0 \leftarrow 1$ subband, all six symmetry components are observed, though the important “1s” $P(1)$ line is mostly obscured by $D_2O$ monomer lines. In the $K_a = 2 \leftarrow 1$ subband, asymmetry doubling is present and sometimes resolved. The “1s” lines are fairly well resolved, but $A_2$ lines appear only as a shoulder on the high side of the stronger $E_2$ transitions, so $A_2$ is naturally less well determined. The fitted parameters for the $K_a$ = 0 and 2 upper states are shown in Table II.

The line widths in Table II were determined using profile fits in which a certain Gaussian profile was assumed for the ‘instrumental’ width (usually about 0.002 $cm^{-1}$ full width at half maximum for He carrier gas) which includes Doppler broadening (due to non-orthogonality of the laser beam and jet propagation) and laser jitter/drift effects. An additional Lorentzian profile was then included for each symmetry ($A_1$, $B_1$, etc.) and $K_a$ value to account for upper state lifetime (predissociation) broadening. The rotational temperature was also varied, and the best fits usually gave values around 1.8 K (for He carrier gas), which was lower than the 2.5 K value which we would otherwise estimate. This discrepancy arises from a non-Boltzmann population distribution, common in supersonic expansions, for which the effective temperature is higher for higher energy levels. That is, the profile fits were dominated by the strong low-$J$ lines (giving lower temperatures), while our usual estimates compared low-$J$ lines with weaker higher-$J$ ones (giving higher temperatures). A number of factors contributed to uncertainty in the width determinations. In addition to temperature effects, the exact contribution of instrumental broadening was somewhat uncertain since, for example, it is likely different for $(D_2O)_2$ than for $D_2O$, and it may not be exactly Gaussian in shape. Moreover, lifetime broadening effects could be $J$-dependent in addition to being symmetry and $K_a$ dependent. Because of these difficulties, we do not give uncertainty estimates for the widths in Table II. Nevertheless, we still believe that these values are meaningful, especially relative to each other, and the excellent fit of the $Q$-branch profile shown in Fig. 3b supports this assertion.

### B. The donor free O-D stretch band

The observed spectrum of this region is shown in Fig. 5a, and we assign parallel ($a$-type) $K_a = 0 \leftarrow 0$ subbands to the "1s" and "2s" states as labeled in black and blue, respectively. Each line of the "1s" band has two components, the lower being assigned to $B_1$ and the upper to a blend of $E_1$ and $A_1$. The three components of the "2s" band are assigned in just the same way, though here there is some ambiguity as discussed below. Due to the extensive blending, all the parameters for this band, as shown in Table III, were determined directly from a profile fit (as opposed to a line position fit). The PGOPHER[24] contour fit option is remarkably powerful and useful, but the resulting parameter error estimates seem to be unrealistically small in the present case, so we believe the statistical uncertainties quoted in Table III underestimate their true values. The simulated profile in Fig. 5a provides a very good representation of the "1s" band. However, in the "2s" band the (already weak) assigned $B_2$ component seems to disappear for $J' >$ 2 or 3, which could be evidence of an upper state perturbation (unfortunately, the relevant region for the $R(2)$ line is obscured by $D_2O$ monomer lines).

In the only previous analysis of this band, Paul et al.[4] assigned a single unresolved $K_a = 0 \leftarrow 0$ subband which agrees well with our "1s" band. In addition, they assigned a single $K_a = 1 \leftarrow 1$ subband,[4] but this coincides almost exactly with our "2s" $K_a = 0 \leftarrow 0$ subband: their $P(2)$ line corresponds to our $P(1)$ line, their $Q$-branch to our $R(0)$, their $R(1)$ to our $R(2)$, etc. Based on our clearly observed $R(1)$ line, which cannot be explained by their assignment, we are confident that the “2s” $K_a = 0 \leftarrow 0$ assignment given here is correct (in spite of the possible perturbation mentioned above). But then the question arises: where are the real $K_a = 1 \leftarrow 1$ subbands for “1s” and “2s”, for which there seems to be no evidence in our spectra? We conclude that they are either hidden under the much stronger $K_a = 0 \leftarrow 0$ bands, or else they remain undetected due to weakness and predissociation broadening.

**C. The donor bound O-D stretch band**

The observed spectrum of the bound O-D stretch fundamental is shown in Fig. 5b. We easily assigned the strongest lines here to the “1s” *a*-type $K_a = 0 \leftarrow 0$ subband, as labeled with black letters. In each line of this band, the $A_1$, $E_1$, and $B_1$ components are just barely resolved, but this is sufficient to make their respective assignments clear thanks to intensity alternation. All other transitions in this region are much weaker. The blue assignments in Fig. 5b indicate what we think must be the “2s”$K_a = 0 \leftarrow 0$ subband, whose weakness is only partly explained by its greater line widths. In this subband, the expected lines with $J' > 3$ ($R(3)$, $P(5)$, etc.) are missing, suggesting the onset of predissociation or some other upper state perturbation.

Somewhat sharper lines, labeled in green below the observed trace in Fig. 5b, can be assigned as a $K_a = 1 \leftarrow 1$ subband with its $Q$-branch at 2632.235 $cm^{-1}$. We think that it is most likely due to the “2s” symmetry component, based on its intensity and position, as explained below. These three labeled subbands in Fig. 5b explain most, but not all, of the observed spectrum. Most remaining features can then be explained in terms of another $K_a = 1 \leftarrow 1$ subband with its $Q$-branch at 2632.14 $cm^{-1}$, about 0.1 $cm^{-1}$ below the green-labeled band. We assign this as the “1s” $K_a = 1 \leftarrow 1$ band with unresolved $A_1$, $E_1$, $B_1$ components. These $K_a = 1 \leftarrow 1$ subband assignments are tentative, and might possibly be interchanged. But they do result in reasonable tunneling splittings, as discussed in Section IV.B. The fitted parameters for the bound

O-D stretch region are listed in Table IV. Previously, Paul et al.[4] assigned a single $K_a = 0 \leftarrow 0$ band in this region which agrees with our "1s" assignment (black in Fig. 5b), and a single $K_a = 1 \leftarrow 1$ band which agrees with our tentative "2s" $K_a = 1 \leftarrow 1$ assignment (green in Fig. 5b). The fact that our "2s" $K_a = 0 \leftarrow 0$ subband (blue in Fig. 5b) seems to disappear for $J' > 3$ could help explain why it was not observed by Paul et al.[4], especially in view of their higher effective rotational temperature (≈10 K).

**D. A weak combination band**

A very weak feature near 2754 cm$^{-1}$, illustrated in Fig. 6, can be assigned as a "1s" subband of $(D_2O)_2$ with $K_a = 1 \leftarrow 0$. This must be normal $(D_2O)_2$, because the line spacing is not compatible with isotopologues containing $^{18}O$, and the triplet structure of each line is not compatible with isotopologues containing H.[13] The $Q$-branch is red-shaded, opposite to those in Fig. 3, suggesting a $c$-type rather than $b$-type transition.

We conclude that this feature is most likely a combination band involving a low-frequency intermolecular mode plus either the donor bound O-D stretch (2632.35 cm$^{-1}$) or the acceptor symmetric stretch (≈2680 cm$^{-1}$). The observed $K_a = 1$ intermolecular state thus has its origin at an energy of 121.8 cm$^{-1}$ relative to the bound O-D stretch, or else at ≈74 cm$^{-1}$ relative to the symmetric stretch origin. The former assignment seems more likely. A fairly recent calculation[17] predicts a value of 122.7 cm$^{-1}$ for the "1s" $K_a = 1$ level of the donor twist overtone vibration, relative to the ground vibrational state (see Fig. 4 of Ref. 17). This would give $c$-type selection rules, as we seem to observe, since the bound O-D stretch ($A'$) plus donor twist overtone ($A'$) combination mode has $A'$ symmetry. Note however, that an earlier calculation[25] assigns this calculated mode as the donor twist fundamental plus acceptor wag, which would give $b$-type selection rules. In either case, the observed $Q$-branch shading might not be definitive since there is ample scope for perturbations in the complicated $(D_2O)_2$ intermolecular vibrational manifold.

The alternative assignment to a combination involving the acceptor symmetric stretch seems less likely, in part because the symmetric stretch fundamental itself is weak and unobserved. There is a $(D_2O)_2$ intermolecular $K_a = 1$ level observed[26] at 68.3 cm$^{-1}$ relative to the ground state, which is not too far from our value of ≈74 cm$^{-1}$. This belongs to the donor twist fundamental mode, and would give *b*-type selection rules in our case, since the acceptor symmetric stretch plus donor twist combination has *A″* symmetry.

Overall, we tentatively assign the 2754 cm$^{-1}$ feature as the "1s" $K_a = 1 \leftarrow 0$ subband of the donor bound O-D stretch plus donor twist overtone combination mode. Each line in Fig. 6 has three components which appear to have approximate predissociation widths of about 0.008 (lower), 0.005 (middle), and 0.004 cm$^{-1}$ (upper component). Assigning their interchange symmetry is difficult due to the poor signal to noise ratio of the spectrum. If we assign the upper, middle, and lower components as $A_1$, $E_1$, and $B_1$, we obtain origin values of 2754.127, 2754.135, and 2754.134 cm$^{-1}$, respectively. This gives a small upper state interchange splitting, 0.008 cm$^{-1}$, which is similar to that observed for the bound O-D stretch fundamental, 0.011 cm$^{-1}$ (Table IV). On the other hand, interchanging these $A_1$ and $B_1$ assignments seems to give better agreement with the observed line intensities, and this assignment is used for the simulation in Fig. 6. This gives origins of 2754.096, 2754.135, and 2754.166 cm$^{-1}$, for $A_1$, $E_1$, and $B_1$, and a larger value, 0.070 cm$^{-1}$, for the interchange splitting which does not agree as well with the fundamental. It is interesting that the transitions in this combination band are significantly sharper (0.004 – 0.008 cm$^{-1}$) than those of the donor bound O-D stretch fundamental itself (0.014 – 0.019 cm$^{-1}$). The observation of this mid-infrared combination band opens up a possible new avenue for investigating water dimer intermolecular vibrations. Note in particular that our favored assignment involves a mode (donor twist overtone) which has not yet been observed directly in the far infrared region.

### E. The acceptor asymmetric O-D stretch bands of $D_2O$ – DOH and $D_2O$ - HOD

A clear but relatively weak subband centered at 2789.0 $cm^{-1}$ (indicated with an asterisk in Fig. 3a) was assigned to the $K_a = 1 \leftarrow 0$ acceptor asymmetric O-D stretch of the deuteron-bound mixed isotopologue, $D_2O$-DOH. This is actually the "1s" component, and a total of 10 *P* and *R* lines were measured in addition to 5 *Q* lines. The corresponding $D_2O$-DOH "2s" subband was also assigned, but it was weaker and its *Q*-branch was partly obscured by $(D_2O)_2$ "1s" *P*(2) transitions at 2787.0 $cm^{-1}$, so only 8 lines could be measured. These assignments are based in part on excellent agreement with the ground state rotational parameters of $D_2O$-DOH as determined by Fraser et al.[13] $D_2O$-DOH has acceptor switching splitting ("1s", "2s"), but no donor acceptor interchange splitting (A, B, E) since the two component monomers ($D_2O$ and DOH) are not identical. Thus only a single component was observed for each "1s" or "2s" *P*, *Q*, or *R* transition, as expected. The parameters resulting from fitting the observed $D_2O$-DOH spectrum are listed in Table V. The $D_2O$-DOH lines were fairly sharp, with estimated predissociation widths of about 0.0025 $cm^{-1}$. Finally, an even weaker band was assigned to the $K_a = 1 \leftarrow 0$ acceptor asymmetric O-D stretch of the mixed proton-bound isotopologue, $D_2O$-HOD, and we presume that it is likely due to the stronger "1s" component. Seven lines were assigned, including the unresolved *Q*-branch at 2787.791 $cm^{-1}$, and the resulting parameters are shown in Table V. We were not able to detect any additional bands for $D_2O$-DOH or $D_2O$-HOD.

## III. Discussion and conclusions

### A. Intensities

From the equilibrium geometry of water dimer (Fig. 1), one might expect that the free O-D stretch fundamental should have mostly *c*-type character, with a much smaller *a*-type component. But in fact the observed band is *a*-type for $(D_2O)_2$, as shown clearly here and in Ref. 4. We were not able to detect *c*-type $K_a = 0 \leftarrow 1$ and $1 \leftarrow 0$ bands, even knowing exactly where to search. In the case of the analogous free O-H stretch in $(H_2O)_2$, Huang and Miller[7] in 1989 originally assigned *c*-type "overlapping $K_a = 0 \leftarrow 1$ and $1 \leftarrow 0$ subbands" around 3732 $cm^{-1}$. But their analysis was not entirely convincing, and more recent observations of this band for $(H_2O)_2$ trapped in helium nanodroplets have been interpreted in terms of *a*-type $K_a = 0 \leftarrow 0$ transitions.[9,17,27]

Our profile fits gave information on the relative intensities of the different symmetry states, subject of course to uncertainties related to temperature and line widths. For the *b*-type acceptor asymmetric stretch subbands, the observed intensities as used for the simulated spectra (Figs. 3 and 4) agree well with the expected ground state populations based on the idea that the "1s" and "2s" states relax separately in the supersonic expansion. That is, they do not thermally interconvert in the time scale of our expansion, so that the "2s" retain most of their 33% population share in spite of their higher energy.

However, the intensities observed for the *a*-type free and bound O-D stretch bands did not always seem to be consistent with these "expected" ground state populations that worked for the asymmetric stretch. We already noted the weakness of the bound stretch "2s" band (Fig. 5b), for which the best simulated spectrum required that the "2s" intensity be reduced to about 40% of its expected value. This weakness could possibly be explained by assuming that the observed lines are due to $A_2$ and/or $B_2$, and that the $E_2$ component is very broad and undetected. For the free stretch (Fig. 5a), the best fit was obtained by using about 65% of the expected "2s" intensity.

A further possible anomaly affected the “1s” states, where the best fits for the bound and free stretch bands seemed to require reduced intensity for $E_1$ relative to $A_1$ and $B_1$, though this was uncertain because the lines were not fully resolved in either case. The ground state populations could not have been significantly different when the spectra of the different bands were recorded, so why should the bound and free O-D stretch bands require different intensity weights than the asymmetric stretch bands? One possibility is that transition moments could perhaps depend on tunneling state. Indeed, calculations[28] have shown quite large differences between the transition moments of“1s” and “2s” states for some water dimer intermolecular modes, and perhaps this could also be possible for the intramolecular modes studied here.

**B. Tunneling splittings**

The value assumed here for the $K = 0$ ground state “1s” to “2s” splitting (53.0 GHz) was in fact determined by Paul et al.[3] on the basis of their original study of the $(D_2O)_2$ acceptor asymmetric O-D stretch band. Our current observation of this band with increased resolution does not significantly affect this determination since it was only approximate at best.

Upper state tunneling splittings can be determined from the origin values listed in Tables II – IV, keeping in mind that all “2s” levels have to be raised (or lowered) depending on whether the true ground state $K_a = 0$ acceptor switching splitting exceeds (or falls short of) the assumed value of 53.0 GHz. Plots in Figs. 7 and 8 help to visualize these splittings and show how they differ from those in the ground state. In all cases, the excited vibrational state acceptor switching splittings (“1s” to “2s”) are modestly reduced, with values ranging from 68% to 86% of their ground state values. This tells us that the effective barrier to the tunneling motion (the “methyl amine like” interchange of acceptor deuterons) becomes slightly higher when any of the O-D stretch fundamentals studied here are excited.

The bound O-D stretch $K_a$ = 1 results from Table IV (not shown in Fig. 8) give an excited state "1s" to "2s" splitting of 0.527 cm$^{-1}$, which is 85% of the known ground state $K_a$ = 1 splitting of 0.619 cm$^{-1}$. This is very similar to the value of 86% observed for $K_a$ = 0 of the same vibration, thus providing support for the present assignment of the two observed $K_a = 1 \leftarrow 1$ subbands. If this assignment were reversed, then the excited state $K_a$ = 1 splitting would be 0.715 cm$^{-1}$, equal to 116% of the ground state value.

More dramatic reductions are observed for the interchange (A to E to B) splittings. This is an expected effect which has been observed previously for intramolecular vibrational bands of water dimer[6,7] and other species, notably HF dimer.[29,30] The idea is that interchanging the roles of the two $D_2O$ monomers in $(D_2O)_2$ becomes more "difficult" in the excited state because the excitation (O-D stretch) then has to be transferred as part of the interchange. More difficult interchange means a higher effective barrier and thus a smaller tunneling splitting. Looking more closely, we find a number of cases in Figs. 7 and 8 where the expected ordering of the A and B levels is inverted, and/or where the E levels are not located between their A and B partners. We conclude that O-D stretch vibrational excitation has strong effects on both the interchange and bifurcation tunneling motions in $(D_2O)_2$.

**C. Line widths, vibrational shifts, and alpha values**

Looking at the measured predissociation widths for the acceptor asymmetric stretch bands in the last column of Table II, we find considerable variation, ranging from the rather sharp $K$ = 1, $E_1$ upper state (<0.001 cm$^{-1}$) to the much broader $K$ = 1, $B_2$ state (≈0.03 cm$^{-1}$). But there is not much apparent regularity in this variation apart from the fact that A states tend to be sharper than B states. For the free and (especially) bound donor O-D stretch bands in Tables III and IV, broader lines are generally observed, with less variation between tunneling states. The larger widths are perhaps not surprising since these modes seem more likely to couple to the dimer dissociative motion than does the acceptor asymmetric stretch.

Averaging over the six tunneling states, the vibrational interval of the $D_2O$ dimer acceptor asymmetric stretch fundamental is about 2783.12 $cm^{-1}$ which represents a modest red shift of about -4.60 $cm^{-1}$ relative to the free $D_2O$ monomer. The donor free and bound O-D stretch fundamentals have average intervals of about 2763.10 and 2632.27 $cm^{-1}$, respectively. These dimer vibrations represent mixtures of the monomer symmetric (2671.645 $cm^{-1}$) and asymmetric (2787.718 $cm^{-1}$) stretches,[31] so we can say that, in some sense, the average shift is about -32 $cm^{-1}$. This negative vibrational shift implies that the dimer is significantly more strongly bound in these excited vibrational states. Of course there is one additional dimer O-D stretch fundamental which we have not observed here, namely the acceptor symmetric stretch which as mentioned lies at 2671.645 $cm^{-1}$ for the monomer. The corresponding dimer band, which could be *a*- and/or *c*-type, is believed to lie around 2680 $cm^{-1}$ and is evidently very weak.[4]

The observed alpha values (changes in $B_{av}$ upon vibrational excitation) are almost all positive. Average alphas are roughly +0.0002 $cm^{-1}$ for the acceptor asymmetric stretch and donor free O-D stretch, and +0.001 $cm^{-1}$ for the donor bound O-D stretch. These increases in $B_{av}$, the opposite of what is usually observed for “normal” molecules, imply tighter bonding in the dimer excited states, especially the bound O-D stretch. This could be related to the deeper potential wells implied by the red shifts mentioned in the previous paragraph. Alphas for the acceptor asymmetric stretch band of $D_2O$-DOH and $D_2O$-HOD are quite similar to those of $(D_2O)_2$ (Table V). For the $H_2O$ dimer O-H stretch fundamentals, the only precise data is for $K = 0$ of the acceptor asymmetric stretch[7] where again a similar small positive alpha was observed relative to the ground[16] state.

**D. Conclusions**

Five bands of the fully deuterated water dimer in the O-D stretch region have been recorded with high resolution (≈0.002 $cm^{-1}$), enabling observation of all six tunneling components in most cases. The results extend, and mostly confirm, previous lower resolution (0.03 $cm^{-1}$) studies of the same bands.[3,4] Another weak feature was tentatively assigned as a $(D_2O)_2$ combination band involving the donor bound O-D stretch intramolecular mode plus two quanta of the intermolecular donor twist mode. The $D_2O$-DOH and $D_2O$-HOD species were also observed, probably the first such mid-infrared spectra of substituted water dimer isotopologues. The dominant $(D_2O)_2$ tunneling splittings (due to "acceptor switching") were found to decrease moderately upon O-D stretch excitation to 68 – 86% of their ground state values. In contrast, the smaller donor-acceptor interchange splittings were found to undergo large and rather irregular decreases upon O-D stretch excitation. Predissociation effects were highly variable, resulting in line widths ranging from less than 0.001 $cm^{-1}$ to 0.03 $cm^{-1}$ (and probably greater), and implying excited state lifetimes in the range of roughly 5 to 0.2 ns. The results paint a detailed picture of $(D_2O)_2$ rotation-tunneling behavior in the excited O-D stretching fundamentals. Similar high resolution results are not possible for $(H_2O)_2$ due to much larger broadening effects in most of its O-H stretch fundamentals.[7,8]

**Acknowledgements**

The financial support of the Natural Sciences and Engineering Research Council of Canada is gratefully acknowledged. We thank K. Esteki for assistance with the measurements.

**References**

[1] A. Mukhopadhyay, W.T.S. Cole, and R.J. Saykally, Chem. Phys. Lett. **633**, 13 (2015).

[2] A. Mukhopadhyay, S.S. Xantheas, and R.J. Saykally, Chem. Phys. Lett. **700**, 163 (2018).

[3] J.B. Paul, R.A. Provencal, and R. J. Saykally, J. Phys. Chem. A **102**, 3279 (1998).

[4] J.B. Paul, R.A. Provencal, C. Chapo, A. Petterson, and R. J. Saykally, J. Chem. Phys. **109**, 10201 (1998).

[5] J.B. Paul, R.A. Provencal, C. Chapo, K. Roth, R. Casaes, and R.J. Saykally, J. Phys. Chem. A **103**, 2972 (1999).

[6] J.T. Stewart and B.J. McCall, J. Phys. Chem. A **117**, 13491 (2013).

[7] Z.S. Huang and R.E. Miller, J. Chem. Phys. **91**, 6613 (1989).

[8] J.B. Paul, C.P. Collier, R.J. Saykally J.J. Scherer, and A. O'Keefe, J. Phys. Chem. A **101**, 5211 (1997).

[9] K. Kuyanov-Prozument, M.Y. Choi, and A.F. Vilesov, J. Chem. Phys. **132**, 014304 (2010).

[10] S.A. Nizkorodov, M. Ziemkiewicz, D.J. Nesbitt, and A.E.W. Knight, J. Chem. Phys. **122,** 194316 (2005).

[11] T. Földes, T. Vanfleteren, and M. Herman, J. Chem. Phys. **141**, 111103 (2014).

[12] N. Suas-David, T. Vanfleteren, T. Földes, S. Kassi, R. Georges, and M. Herman, J. Phys. Chem. A **119**, 10022 (2015).

[13] G.T. Fraser, F.J. Lovas, R.D. Suenram, E.N. Karyakin, A. Grushow, W.A. Burns, and K.R. Leopold, J. Mol. Spectrosc. **181**, 229 (1997).

[14] E.N. Karyakin, G.T. Fraser, and R.D. Suenram, Mol. Phys. **78**, 1179 (1993).

[15] E. Zwart, J.J. ter Meulen, and W.L Meerts, Chem. Phys. Lett. **173**, 115 (1990).

[16] H.A. Harker, F.N. Keutsch, C. Leforestier, Y. Scribano, J.-X. Han, and R.J. Saykally, Mol. Phys. **105**, 497 (2007).

[17] C. Leforestier, K. Szalewicz, and A. van der Avoird, J. Chem. Phys. **137**, 014305 (2012).

[18] L.H. Coudert, F.J. Lovas, R.D. Suenram, and J.T. Hougen, J. Chem. Phys. **87**, 6290 (1987).

[19] J.A. Odutola, T.A. Hu, D. Prinslow, S.E. O'Dell, and T.R. Dyke, J. Chem. Phys. **88**, 5352 (1988).

[20] F.N. Keutsch, N. Goldman, E.N. Karyakin, H.A. Harker, M.E. Sanz, C. Leforestier, and R.J. Saykally, Faraday Discussions **118**, 79 (2001).

[21] M. Dehghany, M. Afshari, Z. Abusara, C. Van Eck, and N. Moazzen-Ahmadi, J. Mol. Spectrosc. **247**, 123 (2008).

[22] M. Rezaei, K.H. Michaelian, and N. Moazzen-Ahmadi, J. Chem. Phys. **136**, 124308 (2012).

[23] M. Rezaei, S. Sheybani-Deloui, N. Moazzen-Ahmadi, K.H. Michaelian, and A.R.W. McKellar, J. Phys. Chem. A **117**, 9612 (2013).

[24] C.M. Western, PGOPHER, a program for simulating rotational structure version 8.0, 2014, University of Bristol Research Data Repository, doi:10.5523/bris.huflggvpcuc1zvliqed497r2

[25] H.A. Harker, F.N. Keutsch, C. Leforestier, Y. Scribano, J.-X. Han, and R.J. Saykally, Mol. Phys. **105**, 513 (2007).

[26] L.B. Braly, J.D. Cruzan, K. Liu, R.S. Fellers, and R.J. Saykally, J. Chem. Phys. **112**, 10293 (2000).

[27] R. Fröchtenicht, M. Kaloudis, M. Koch, and F. Huisken, J. Chem. Phys. **105**, 6128 (1996).

[28] See Table 13 of: M.J. Smit, G.C. Groenenboom, P.E.S. Wormer, A. van der Avoird, R. Bukowski, and K. Szalewicz, J. Phys. Chem. A **105**, 6212 (2001).

[29] A.S. Pine and W.J. Lafferty, J. Chem. Phys. **78**, 2154 (1983).

[30] I.M. Mills, J. Phys. Chem. **88**, 532 (1983).

[31] R.A. Toth, J. Mol. Spectrosc. **195**, 98 (1999).

Table I. Symmetry labels and assumed ground state origins[16] for $(D_2O)_2$ (in $cm^{-1}$).

| Present notation | Complete label | $K$ = 0 origin | $K$ = 1 origin |
|---|---|---|---|
| $A_1$ | $A_1^+/B_1^-$ | 0.00000 | 5.36232 |
| $E_1$ | $E_1^+/E_1^-$ | 0.02002 | 5.37980 |
| $B_1$ | $B_1^+/A_1^-$ | 0.03910 | 5.39822 |
| $A_2$ | $A_2^-/B_2^+$ | 1.76937 | 4.74511 |
| $E_2$ | $E_2^-/E_2^+$ | 1.78791 | 4.76119 |
| $B_2$ | $B_2^-/A_2^+$ | 1.80551 | 4.77821 |

Table II. Upper state parameters for the acceptor asymmetric stretch fundamental band of $(D_2O)_2$ (in $cm^{-1}$). [a]

| $K_a$ | Sym | Origin, σ | $B_{av}$ | $10^6 \times D$ | $10^4 \times (B - C)/4$ | Width |
|---|---|---|---|---|---|---|
| 0 | $A_1$ | 2784.7910(6) | 0.181490(92) | [1.2] | | 0.0024 |
| 0 | $E_1$ | 2784.7880(6) | 0.181468(75) | [1.2] | | 0.0048 |
| 0 | $B_1$ | 2784.7871(6) | 0.181541(92) | [1.2] | | 0.0097 |
| 0 | $A_2$ | 2783.2601(5) | 0.181435(64) | [1.2] | | 0.003 |
| 0 | $E_2$ | 2783.2613(3) | 0.181383(18) | [1.2] | | 0.003 |
| 0 | $B_2$ | 2783.2629(4) | 0.181342(42) | [1.2] | | 0.003 |
| 1 | $A_1$ | 2787.9215(1) | 0.181322(8) | 1.24(13) | 4.595(17) | 0.0015[b] |
| 1 | $E_1$ | 2787.9241(1) | 0.181319(8) | 1.19(10) | 4.591(22) | 0.0004[b] |
| 1 | $B_1$ | 2787.9263(1) | 0.181309(11) | 1.04(11) | 4.516(49) | 0.0029[b] |
| 1 | $A_2$ | 2788.3939(1) | 0.181350(20) | 0.44(46) | 3.0(15) [c] | 0.0014[b] |
| 1 | $E_2$ | 2788.3913(1) | 0.181472(14) | 2.39(24) | 3.0(15) [c] | 0.0041[b] |
| 1 | $B_2$ | [2788.399] | [0.18132] | [1.2] | [3.0] | [0.03][b] |
| 2 | $A_1$ | 2800.0721(7) | 0.181204(27) | [1.2] | | 0.0023 |
| 2 | $E_1$ | 2800.0706(7) | 0.181243(27) | [1.2] | | 0.0032 |
| 2 | $B_1$ | 2800.0742(7) | 0.181184(28) | [1.2] | | 0.0053 |
| 2 | $A_2$ | 2801.107(1) | [0.18111] | [1.2] | | [0.007] |
| 2 | $E_2$ | 2801.1160(11) | 0.181075(77) | [1.2] | | 0.0073 |
| 2 | $B_2$ | 2801.107(1) | [0.18111] | [1.2] | | 0.0090 |

[a] Uncertainties in parentheses are 1σ in units of the last quoted decimal. Quantities in square brackets were held fixed.

[b] Width values for $K_a$ = 1 apply to *R*- and *P*-branches, that is to upper state levels with $J_{KaKc}$ = $1_{11}$, $2_{12}$, $3_{13}$, etc. For the"1s" *Q*-branch, involving $1_{10}$, $2_{11}$, $3_{12}$, etc., we obtain slightly different values of 0.0014, 0.0006, and 0.0020 $cm^{-1}$, respectively (Fig. 3b).

[c] Manually fit to *Q*-branch profile.

Table III. Upper state parameters for the free O-D stretch fundamental of $(D_2O)_2$ (in $cm^{-1}$). [a]

| $K_a$ | Sym | Origin, σ | $B_{av}$ | Width |
|---|---|---|---|---|
| 0 | $A_1$ | 2763.3996(1) | 0.181518(4) | 0.012 |
| 0 | $E_1$ | 2763.4130(3) | 0.181283(8) | 0.020 |
| 0 | $B_1$ | 2763.4035(1) | 0.181392(5) | 0.014 |
| 0 | $A_2$ | 2764.6083(2) | [0.181133] | [0.012] |
| 0 | $E_2$ | 2764.6200(1) | 0.181133(7) | 0.015 |
| 0 | $B_2$ | 2764.5939(2) | [0.181133] | 0.014 [b] |

[a] Uncertainties in parentheses are 1σ in units of the last quoted decimal from a profile fit; true uncertainties are larger. Quantities in square brackets were held fixed. Upper state *D*-values were fixed at $1.2 \times 10^{-6}$.

[b] Estimated manually.

Table IV. Upper state parameters for the bound O-D stretch fundamental of $(D_2O)_2$ (in $cm^{-1}$). [a]

| $K_a$ | Sym | Origin, σ | $B_{av}$ | Width |
|---|---|---|---|---|
| 0 | $A_1$ | 2632.3446(13) | 0.182211(69) | 0.018 |
| 0 | $E_1$ | 2632.3533(14) | 0.182130(100) | 0.014 |
| 0 | $B_1$ | 2632.3556(13) | 0.182258(69) | 0.019 |
| 0 | "2s" | 2633.8672(18)[b] | 0.183128(300) | 0.037 |
| 1 | "1s" | 2637.520(3)[b] | 0.18189(62) | 0.03 |
| 1 | "2s" | 2636.9935(11)[b] | 0.181988(80) | 0.017 |

[a] Uncertainties in parentheses are 1σ in units of the last quoted decimal from a profile fit; true uncertainties are larger. Upper state *D*-values were fixed at $1.2 \times 10^{-6}$.

[b] Here we assume that the lower state is the E component.

Table V. Parameters for the acceptor asymmetric stretch fundamental $K = 1 \leftarrow 0$ subbands of $D_2O$ – DOH and $D_2O$ – HOD (in $cm^{-1}$). [a]

| | Ground state, $K = 0$ [b] | | Excited state, $K = 1$ | | |
|---|---|---|---|---|---|
| | $B_{av}$ | $10^6 \times D$ | Origin | $B_{av}$ | $10^4 \times (B - C)/4$ |
| $D_2O$-DOH “1s” | [0.18929630] | [1.44] | 2789.1814(1) | 0.189509(8) | 9.739(46) |
| $D_2O$-DOH “2s” | [0.18922071] | [1.15] | 2787.1758(2)[c] | 0.189664(12) | 6.42(8) |
| $D_2O$-HOD “1s” | [0.18129365] | [1.22] | 2787.9676(5) | 0.181482(48) | [4.0] |

[a] Uncertainties in parentheses are $1\sigma$ in units of the last quoted decimal.

[b] Ground state parameters were fixed at these values from Fraser et al.[13] Excited state *D* values were fixed at the ground state values.

[c] This is a band origin. To obtain the true “2s” $K = 1$ state origin, the (unknown) energy of the ground “2s” $K = 0$ state relative to $K = 0$ “1s” should be added to this value.

**Figure Captions**

Fig. 1. Equilibrium structure of the water dimer.

Fig. 2. Ground state rotational origins for $(D_2O)_2$, from Harker et al.[16] The six ground state ($J$ = 0, $K_a$ = 0) tunneling components are shown in (a), with spin statistical weights in parentheses. All of the $(D_2O)_2$ spectra reported in this paper originate from the "1s" and "2s" $K_a$ = 0 and 1 states shown in (b).

Fig. 3. Observed (black) and simulated (colors) spectra of the $(D_2O)_2$ acceptor asymmetric O-D stretch $K_a = 1 \leftarrow 0$ subband. An asterisk marks the "1s" $Q$-branch for $D_2O$-DOH. The $(D_2O)_2$"1s" $Q$-branch region shown in (b) was recorded using argon carrier gas, giving an effective rotational temperature of about 4 K and a Gaussian Doppler instrumental width of about 0.0015 $cm^{-1}$. All other spectra shown in this paper used helium carrier, giving temperatures of about 2 K and slightly broader instrumental widths around 0.002 $cm^{-1}$ (see Sec. III.A).

Fig. 4. Observed (black) and simulated (red) spectra of the $(D_2O)_2$ acceptor asymmetric O-D stretch subbands with: (a) $K_a = 0 \leftarrow 1$, and (b) $K_a = 2 \leftarrow 1$.

Fig. 5. Observed (black) and simulated (red) spectra of: (a) the $(D_2O)_2$ donor free O-D stretch band, and (b) the $(D_2O)_2$ donor bound O-D stretch band. Black (or blue) letters label "1s" (or "2s") transitions with $K_a = 0 \leftarrow 0$. Green letters label "2s" transitions with $K_a = 1 \leftarrow 1$.

Fig. 6. Observed (black) and simulated (red) spectra of a weak $(D_2O)_2$ combination band tentatively assigned as due to the donor bound O-D stretch plus the donor twist overtone.

Fig. 7. Comparison of tunneling splittings of $(D_2O)_2$ in the ground vibrational state (left) and the excited acceptor asymmetric O-D stretch state (right). Note the much reduced donor-

acceptor interchange splittings (A, E, B) in the excited state. The excited state $K_a$ = 1, $B_2$ and $K_a$ = 2, $A_2$ levels are less certain than the others.

Fig. 8. Comparison of $K_a$ = 0 tunneling splitting of $(D_2O)_2$ in the ground state (left) and the excited donor free and bound O-D stretch states (middle and right).The free O-D stretch $E_1$ and $A_2$ levels are less certain than the others.

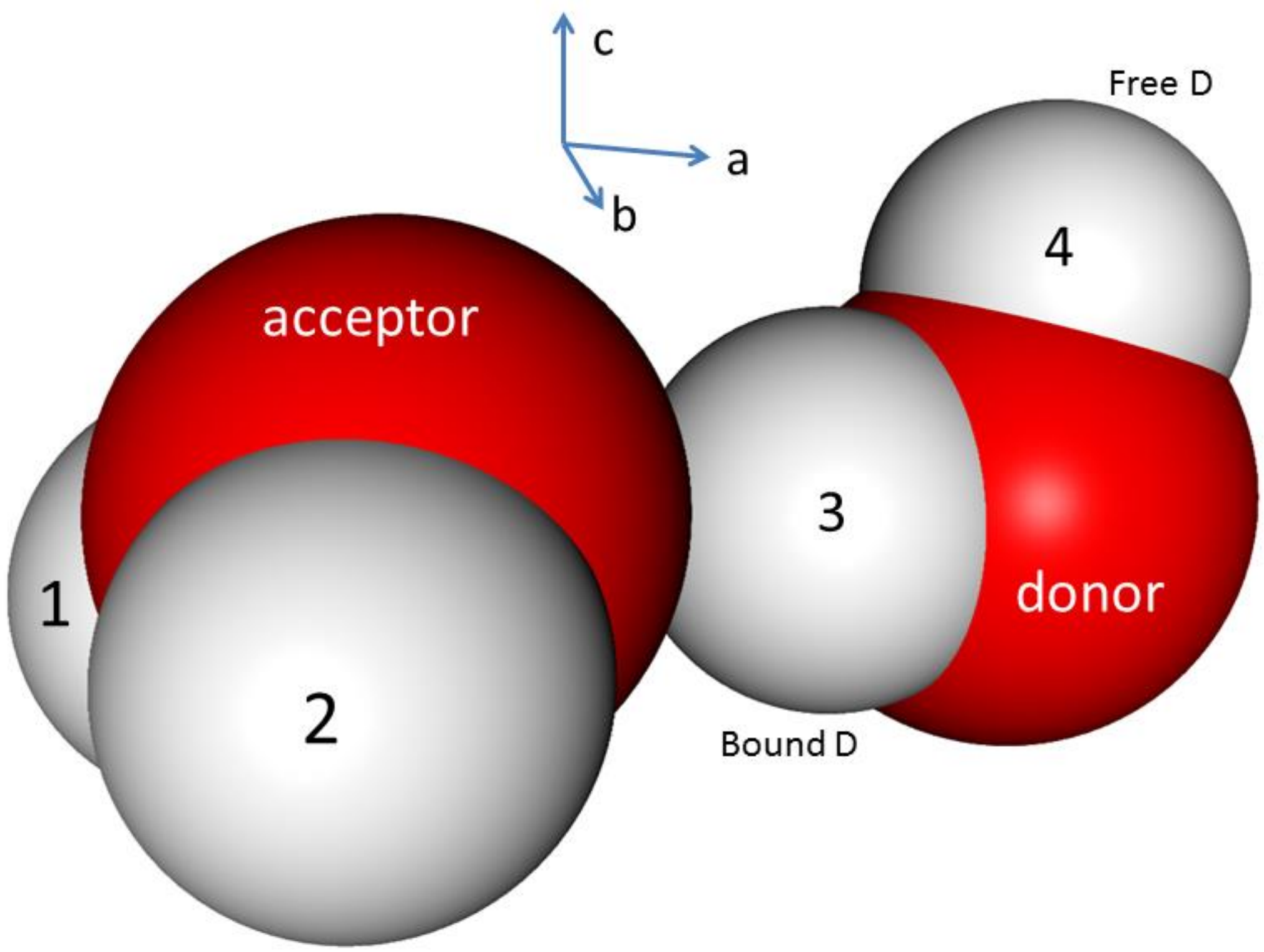

Fig. 1

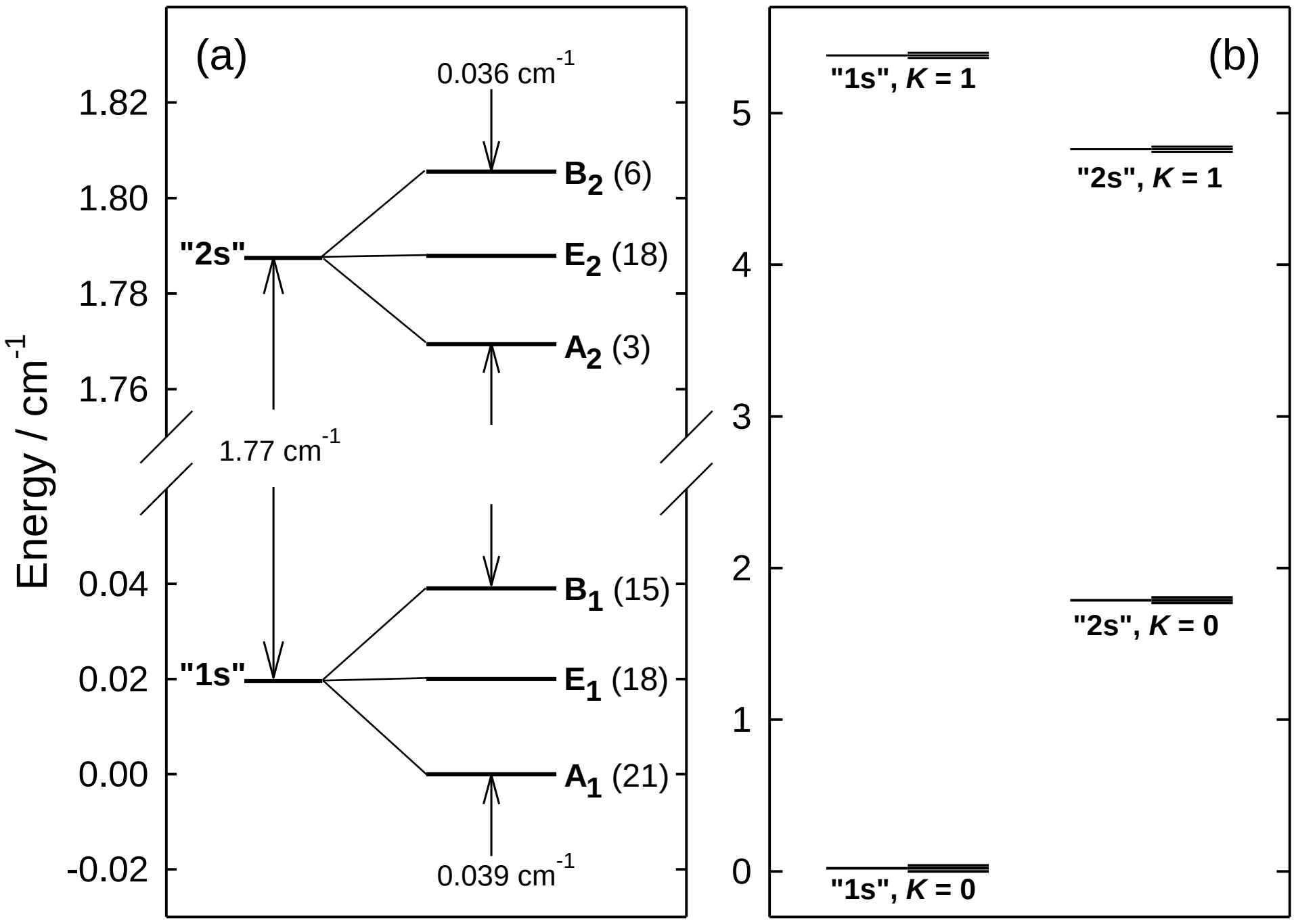

Fig. 2

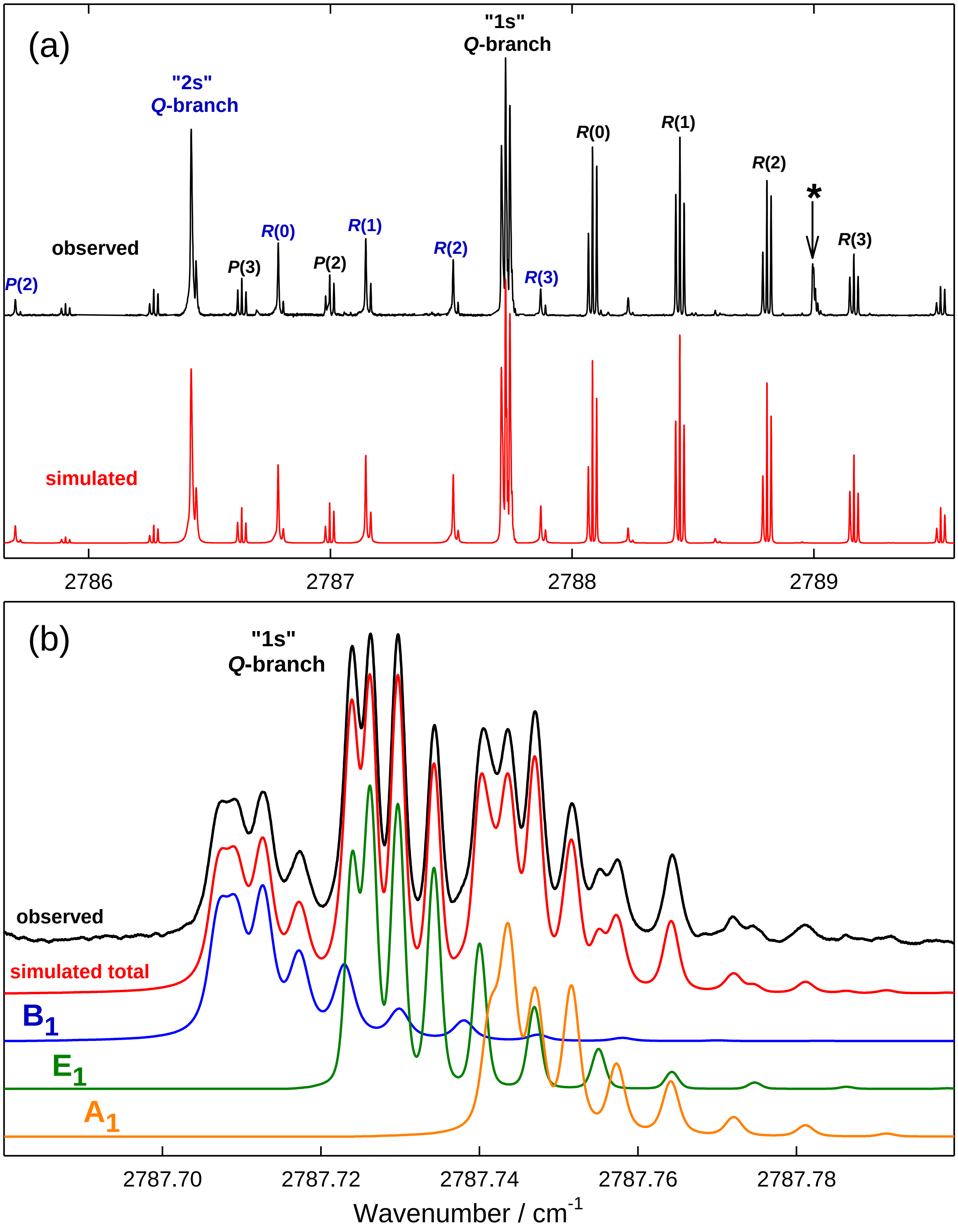

Fig. 3

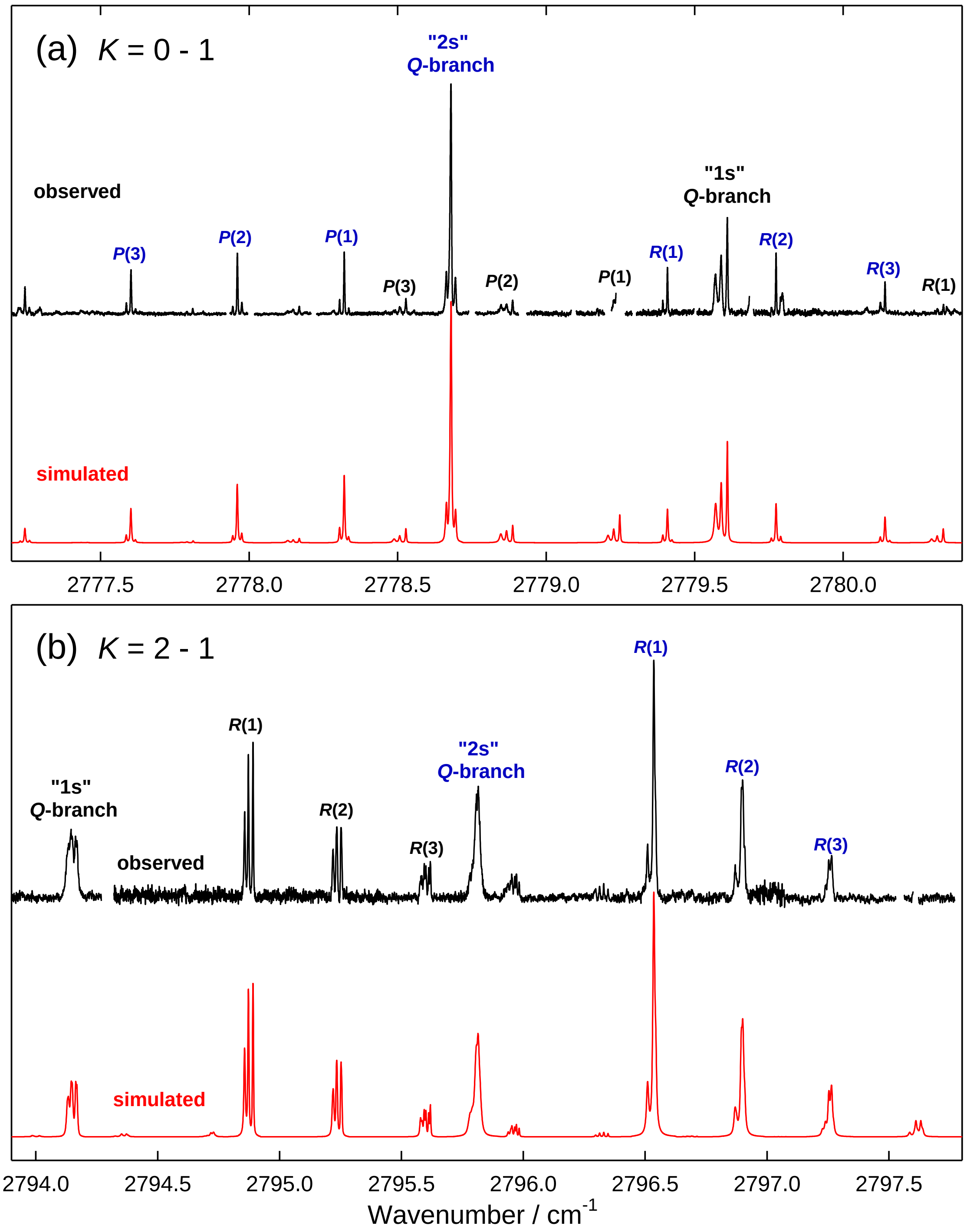

Fig. 4

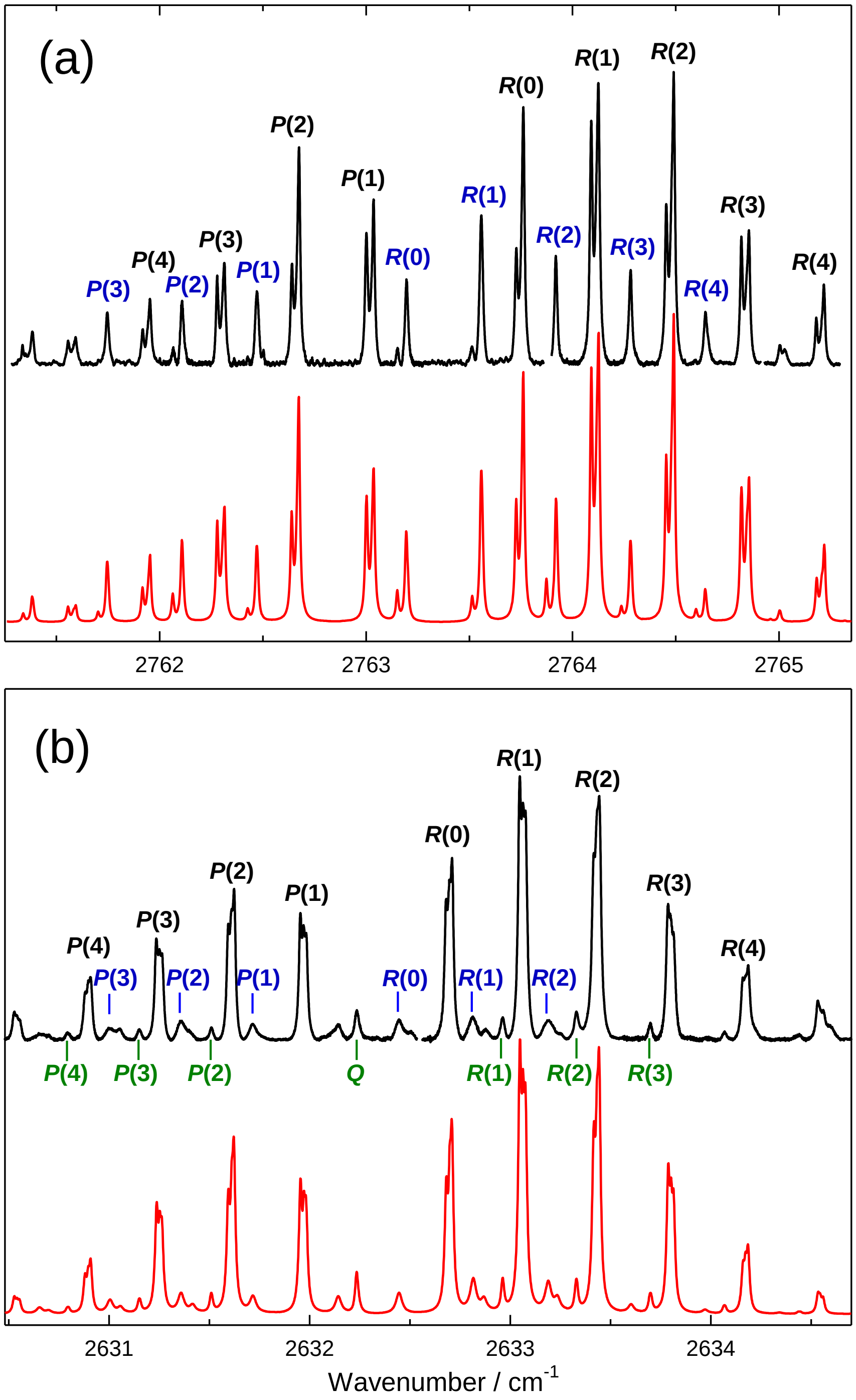

Fig. 5

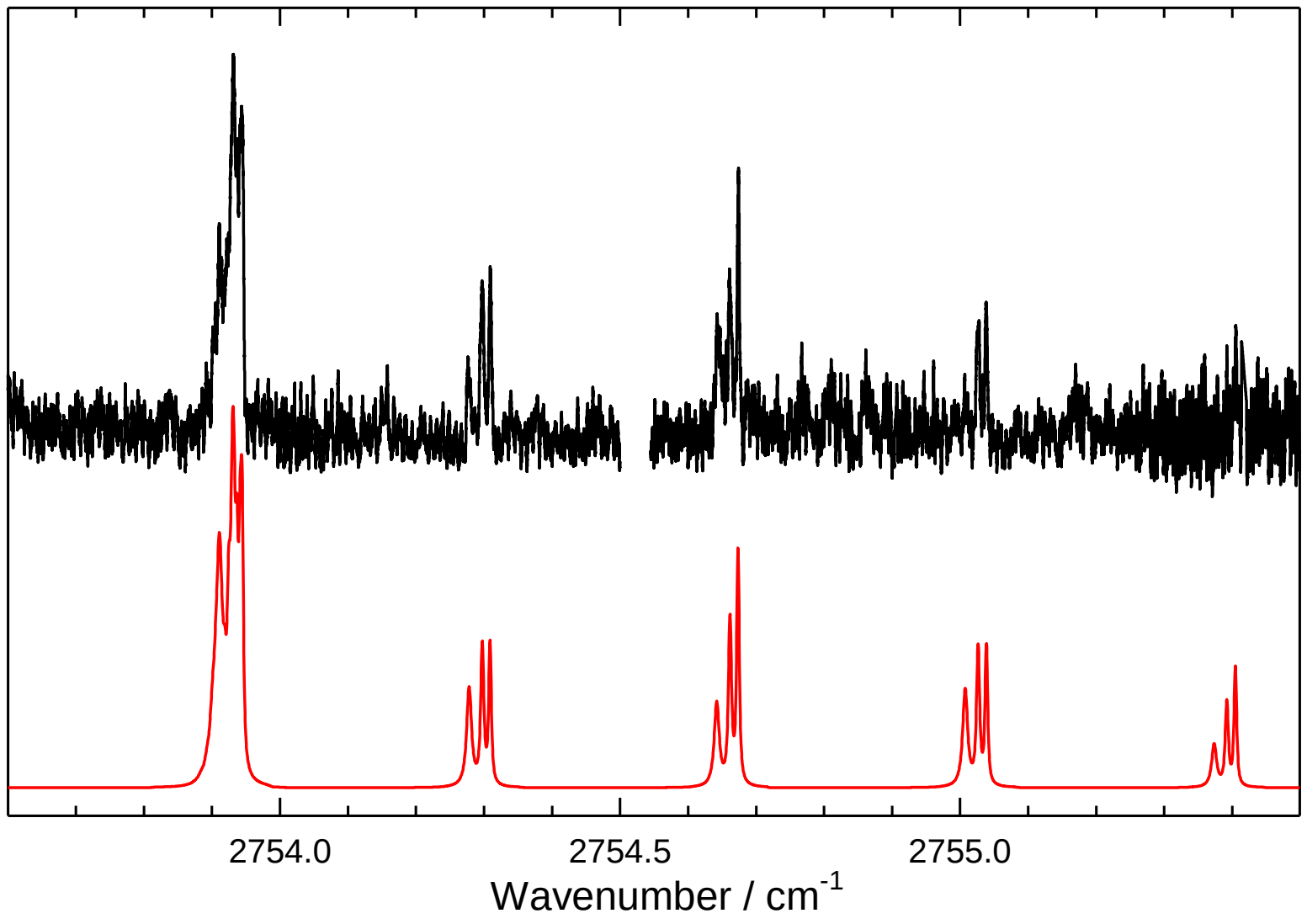

Fig. 6

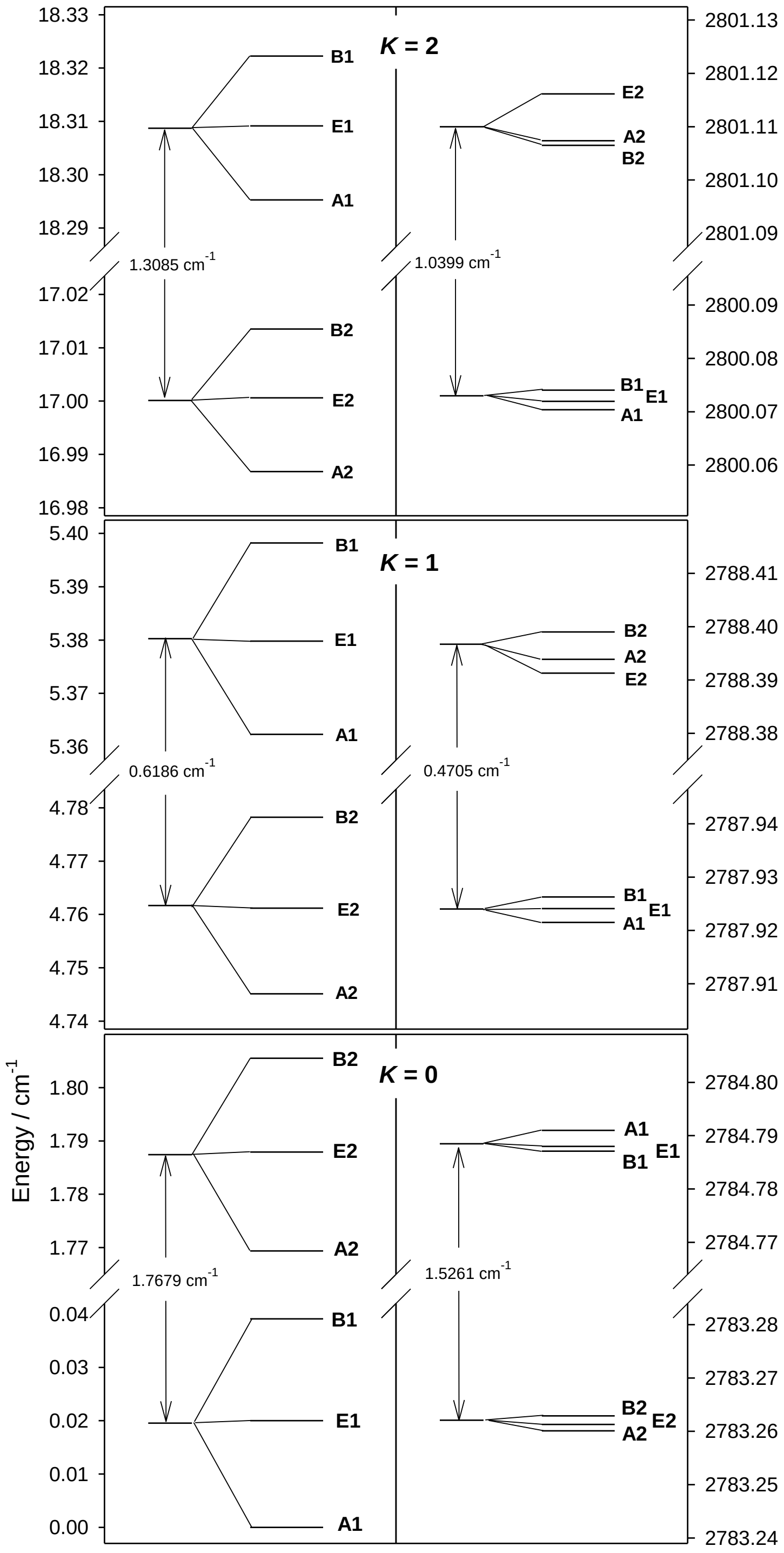

Fig. 7

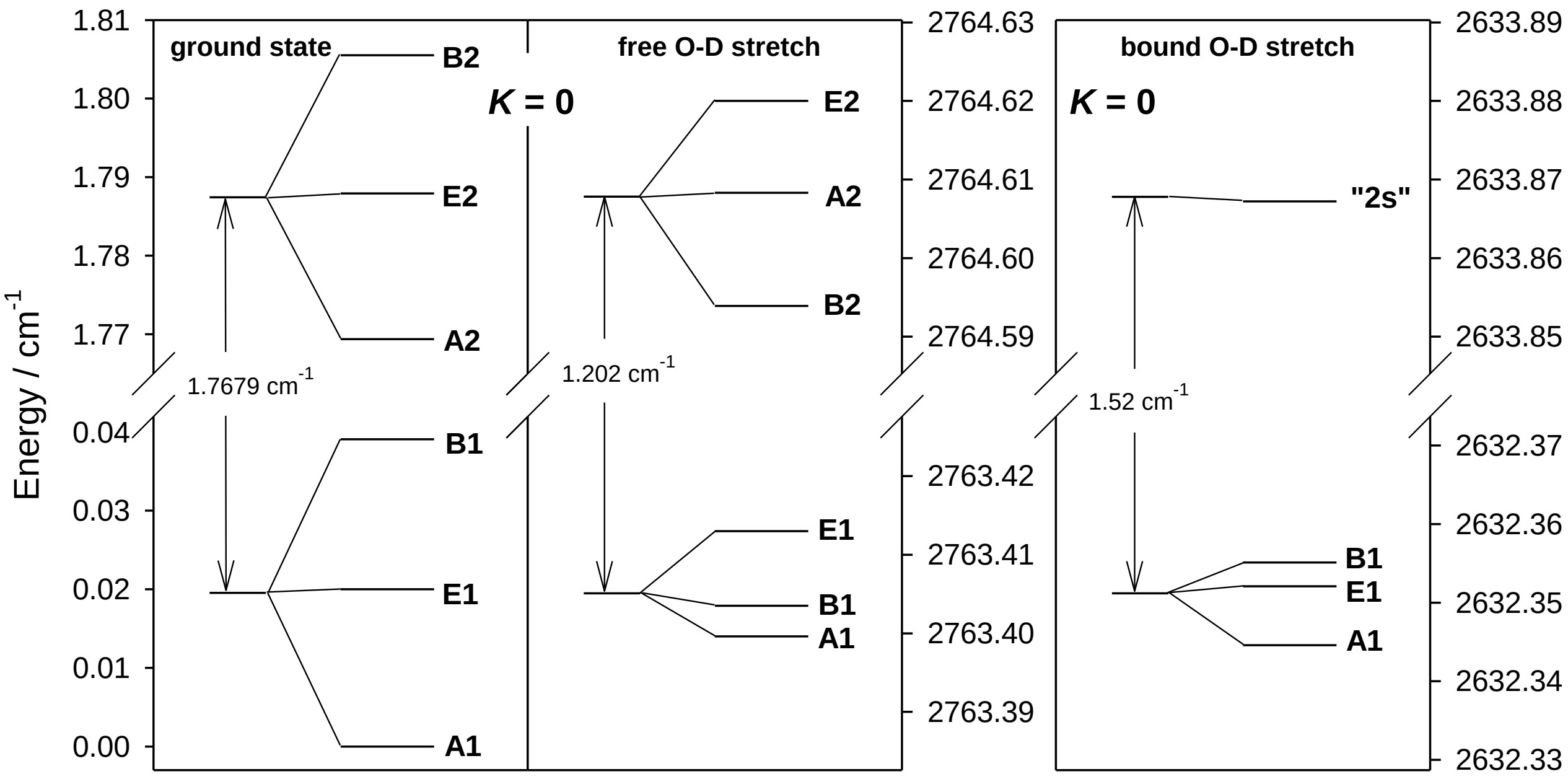

Fig. 8.